\documentclass[pre,twocolumn,showpacs,superscriptaddress] {revtex4}
\usepackage{graphics}
\newcommand{\cu}
{\affiliation{Department of Physics, University of Calcutta,
92 Acharya Prafulla Chandra Road, Kolkata 700009, India}}

\begin{document}
  
 \title
 {Effect of the nature of randomness on quenching dynamics of Ising model on complex networks
}

 \author
 { Soham Biswas}
 \email{soham.physics@gmail.com}
\cu
\author	{  Parongama Sen }
 \email{psphy@caluniv.ac.in}
\cu

\begin{abstract}
Randomness is known to affect the dynamical behaviour of many systems to a large extent. 
In this paper we investigate how the nature of randomness affects the
dynamics in a zero temperature quench of Ising model on two types of random networks. 
In both the networks, which are embedded in a one dimensional space, the first neighbour connections exist
and the average degree is four per node.  
In the random model A,
the second neighbour connections are rewired with a probability $p$ while in the random model B,
additional connections between neighbours at Euclidean distance $l ~ (l >1)$  are introduced with 
a probability $P(l) \propto l^{-\alpha}$. We find that for both models, 
the dynamics leads to freezing such that the system gets locked in a disordered state. The point at which
the disorder of the nonequilibrium steady state is maximum is located. 
Behaviour of  dynamical quantities like residual energy, 
order parameter and persistence are
discussed and compared. 
Overall, the behaviour of physical quantities are similar although subtle 
differences are observed due to the difference in the nature of randomness.

\vskip 0.5cm

\end{abstract}
\maketitle

\section{Introduction}

The dynamics of Ising models is a much studied phenomenon and has emerged as a rich field of present day research.
An important dynamical feature commonly studied is the quenching
phenomenon below the critical temperature.
In a quenching process, the system has a disordered initial configuration
corresponding to a high temperature and its temperature is suddenly dropped.
This results in quite a few interesting phenomena like domain growth \cite{gunton,bray},
persistence \cite{derrida,stauffer,krap1,Krap_Redner} etc.

In one dimension, a zero temperature quench  of the  Ising model 
starting with completely random configuration (which corresponds to a 
very high temperature) and evolving according to the usual Glauber dynamics,
always leads the system to  the equilibrium configuration (all spins up or all spins down).
The average domain size $D$ increases in time $t$ as $D(t)\sim t^{1/z}$,
where $z$ is the dynamical exponent associated with the growth.
As the system coarsens, the magnetisation also
grows in time as $m(t)\sim t^{1/2z}$.
In two or higher dimensions, however, the system does not
always reach equilibrium \cite{Krap_Redner} although these scaling relations
still hold good. In zero temperature quench, another important dynamical behavior commonly studied is persistence,
which is the probability that a spin has not flipped till time $t$. In regular lattices, in one or higher dimensions, 
the persistence probability $P(t)$ at time $t$ is usually seen to follow a power law decay given by $P(t) \propto t^{-\theta}$. 
$\theta$ is called the persistence exponent and
is unrelated to any other known static or dynamic exponents.

The dynamical behaviour of Ising models may change drastically when randomness is introduced in the
system.
Randomness can occur in many ways and its effect on dynamics 
can depend on   its precise nature. For example, randomness
in the Ising model can be incorporated  by introducing dilution in the site or bond occupancy in 
regular lattices and consequently the percolation transition plays an important role \cite{stinch,jain1}. 
Scaling behaviour is completely different from power laws here.
One can also consider the interactions to be randomly distributed, 
either  all ferromagnetic type or mixed type (e.g., as in a spin glass) \cite{stein};  
the system goes to a frozen state following a zero temperature quenching in both cases.
Another way to introduce randomness is to consider a random field 
in which case the scaling behaviour is also completely different from power laws \cite{fisher}.

Here we consider Ising models on random graphs or networks where  
the nearest neighbour connections exist. In addition,  
the  spins have random  long  range interactions which are quenched in nature.
In general, here, the dynamics, instead of leading the system towards its equilibrium state, 
makes it freeze into a metastable state such that the dynamical quantities
attain  saturation values different from their equilibrium values. 

Moreover,  rather than showing a conventional power law decay 
or growth, the dynamical quantities exhibit completely different behaviour in time.

A point to be noted here is, when 
long range links are introduced, the domains are no longer well-defined as interacting neighbours 
could be  well separated in space.  
This results  in freezing of Ising spins on random graphs as well as on small
world networks \cite{sven,haggstrom}.   
The phase ordering dynamics of the Ising model on a Watts-Strogatz network \cite{ws}, 
after a quench to zero
temperature, produces dynamically frozen configurations, disordered at large length scales \cite{boyer,haggstrom}.
Even on small world  networks, the dynamics can  depend on the nature of the randomness;  it was observed that 
while in a sparse network there is freezing, in a densely connected 
network freezing disappears in the thermodynamic limit \cite{pratap_net}.

In this paper, we investigate the dynamical behaviour of an Ising system
on  two different networks  following a zero temperature quench. 
In these two networks, both of which are sparsely connected, 
the nature of randomness is subtly different and
we study whether this difference  has any effect   
on the dynamics. Both these networks are embedded in a one dimensional 
lattice and the nearest neighbour connections always exist and the nodes have degree four on an average. 
They differ as in one of the networks, the random long range interactions
have a spatial dependence. It may be mentioned here that quenching dynamics on such 
Euclidean networks has not been considered earlier to the best of our knowledge. 

It is also quite well known that many dynamical social phenomena 
can be appropriately mapped to dynamics of spin systems. 
 At the same time,  social systems have been shown to  behave like complex
networks (having small world and/or scale free features etc.).   
So the present study may be particularly interesting in the context of  
studying social phenomena described by Ising-type models.

In section II we have described the two different networks 
 which we call  random model A (RMA) and random model B (RMB).
In Section III we have given a list of the quantities calculated.
In section IV and V we have discussed the detailed dynamical behaviour of Ising spin systems on random model A and random model B respectively.
The comparison of the results of the quenching dynamics  between the two models are discussed in section VI. In addition, a qualitative analysis of the 
quenching dynamics is also presented.
Summary and concluding statements are made in the last
section.

\section{Description of the network models}

\begin{figure}[h]
{\resizebox*{6cm}{!}{\includegraphics{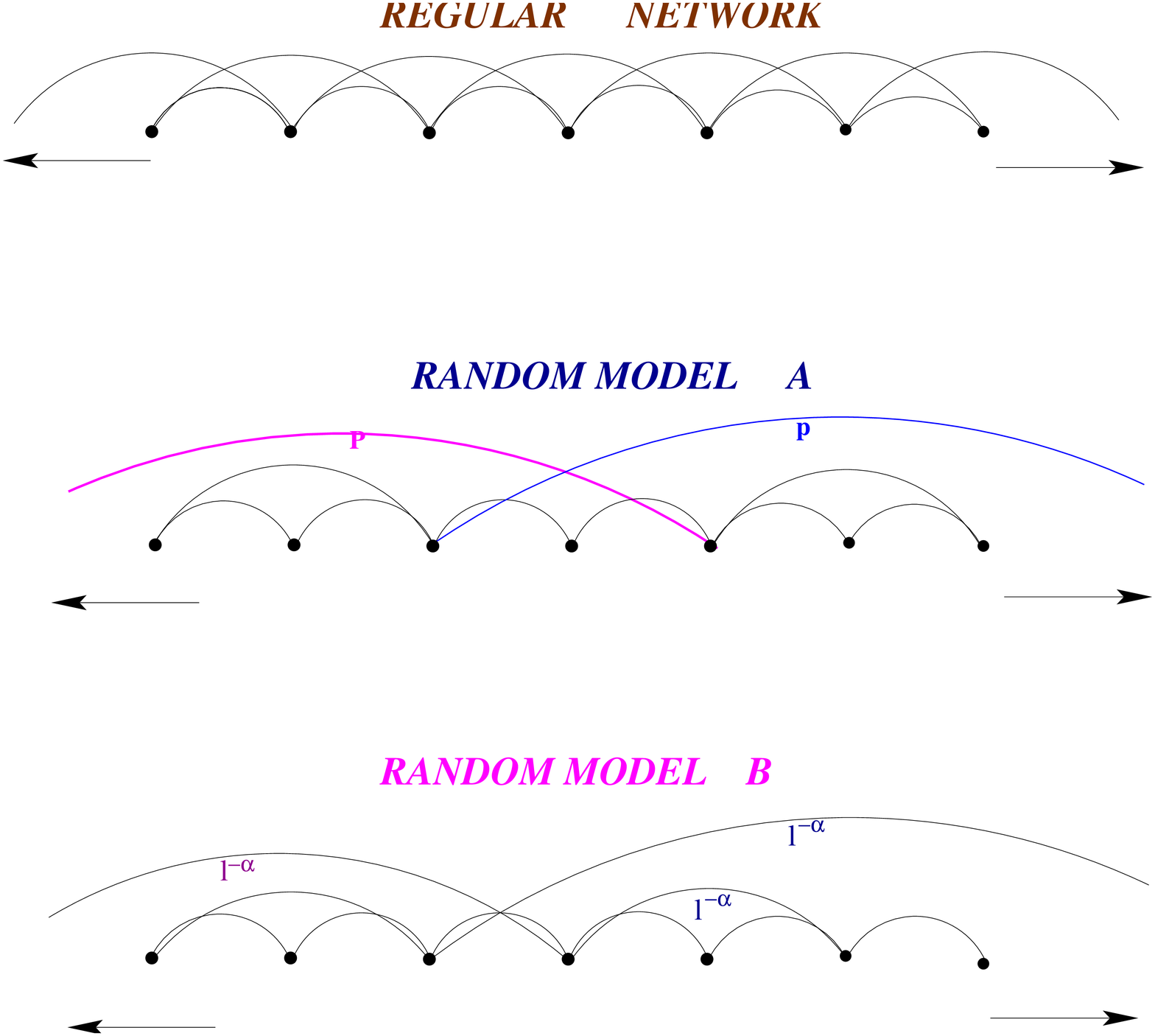}}}
\caption{ (Color online) Schematic diagram for different network models. Average degree is 2K = 4 in each network. 
In the regular network both the first and second nearest neighbours are present. 
In random model A only second neighbours are rewired with probability p.
In random model B first nearest neighbours are always linked while
other nodes are linked with the probability $l^{-\alpha}$ with $ l \geq 2 $.}  
\label{schematic}
\end{figure}

The two network models under consideration were introduced in reference \cite{gbs}.
The random model A  (RMA) is in fact very similar to the Watts-Strogatz network \cite{ws}. Here  initially 
a spin is connected to its four nearest neighbours 
 and then only the second nearest neighbour links are rewired  with probability $p$ (Fig. \ref{schematic}).
 In the RMB, each spin is connected to its  two  nearest neighbour and then
two extra bonds (on an average) are attached randomly to each spin. 
The extra bonds
are attached to spins located at a distance $l>1$ with probability $P(l) 
\propto l^{-\alpha}$ (Fig. \ref{schematic}). 
We keep the first neigbours intact in both cases 
to ensure that the networks are connected. Average degree per node is four in both the networks.
The dynamical evolution is considered on the static networks after the process of rewiring/addition of links
is completed.

The general form of the Hamiltonian in a one dimensional Ising spin system 
for RMA and RMB can be written as
\begin{equation}
 H = -  \sum_{i<j} J_{ij}S_i S_j,
 \end{equation}
where $S_i = \pm 1$ and $J_{ij}=J$ when sites $i$ and $j$ are connected and  zero otherwise. (We take $J=1$ in this paper.)
The ground state (minimum energy state at zero temperature) of the Ising spin system
 in  both RMA and RMB is a state with all spins up or all spins down.

 RMA is a variant of WS model with identical static properties.
It is regular for $p=0$, random for $p=1$, and for any $p > 0$, 
the nature of RMA is small world like \cite{ws,gbs}. Euclidean models of RMB type have been studied in a 
few earlier works \cite{euclid1,euclid2,gbs}.  
While it is more or less agreed that 
for $\alpha \leq 1$, the network is  random and  for $\alpha > 2$, it
behaves as a regular network, the nature of the network for intermediate values of 
$\alpha$ is not very well understood. 
According to the earlier studies \cite{euclid1,euclid2,gbs}, it may either have a small world 
characteristic or behave like a finite dimensional lattice. 
In the present work, we assume that RMB has random nature for $\alpha <1$ and for $1<\alpha<2$, it is small world like
 (at least for the system sizes considered here) following the results of \cite{gbs}, which is  based on exact numerical evaluation of shortest distance and clustering
coefficients.
This is also because the Euclidean model considered in \cite{gbs} is exactly
identical to RMB with average degree four, while the average degree  of the Euclidean models 
considered in the other earlier studies is not necessarily equal to four.

In case of RMA, the network is regular and random for only two extreme values $p=0$ and $p=1$ respectively, 
whereas for RMB, the random and regular behaviour of the network are observed over an  extended region.
The regular network corresponding to these two models is the one dimensional Ising spin system
 with nearest neighbour and next nearest neighbour interactions.
We have studied the zero temperature quenching dynamics for this 
 model  also, and the results for the dynamics are identical to that of the 
nearest neighbour Ising spin model.
So it will be interesting to note  how the dynamics is affected by the introduction of randomness 
in the Ising spin system and also how  the difference  in the nature of randomness
of the two models RMA and RMB shows up in the dynamics.

In the simulations, 
single spin flip Glauber dynamics is
used in both cases, the spins are oriented randomly in the initial  state.
We have taken one dimensional lattices of size $L$ with $100 \leq L \leq 1500$ to study 
the dynamics. 
The results are averaged over (a) different initial configurations and
(b) different network configurations. For each system size the number of networks considered is fifty 
and for each network the number of initial configuration is also fifty.
Periodic boundary condition has been used.

\section{Quantities Calculated}
We have estimated the following quantities in the present work.

\begin{enumerate}
\item Magnetisation $m(t)$:  For a Ising spin system with regular connections and having only the ferromagnetic interaction, 
the order parameter is usually the magnetisation, $m= \frac{|\sum_i S_i|}{L} $. $L $ is the size of the system. Magnetisation  
can be considered as the order parameter, even when the connections are random. We have calculated the growth of 
magnetisation with time and also the variation of the saturation value of the magnetisation, $m_{sat}$, with $p$ and $\alpha$
for RMA and RMB respectively.

\item Persistence probability $P(t)$: As already mentioned, 
this is the probability that a spin
 does not flip till time $t$. 

\item Energy $E(t)$: In these networks, domain wall measurement is not very
significant, as domains are ill-defined. 
The presence of domain walls in regular lattices causes an energy cost \cite{boyer}.
 So instead of the number of domain walls,
the appropriate measure for disorder is the residual energy per spin
 $\varepsilon = E - E_0 = E + 4$, where
$E_0=-4 $ is the known ground state energy per spin and $E$ is the 
energy of the dynamically evolving  state.
In fact, the magnetisation is not  a good measure of the disorder either, since even
when the energy is close to the ground state, magnetisation
may be very close to zero (this is also true for the models without randomness). So residual energy measurement is the best way 
to find out whether the system has reached the equilibrium ground state or 
it is stuck in a higher energy nonequilibrium steady state.
We have measured the decay of residual energy $\varepsilon$ with time and the variation of 
its saturation value, $\varepsilon_{sat}$, with $p$ and $\alpha$ for RMA and RMB respectively.

\item Freezing probability: The probability with  which any configuration freezes, i.e.,  
does not reach the ground state (the state with
magnetisation $m=1$ or the state with  zero residual energy) is defined as the 
freezing probability. 

\item Saturation time : It is the time taken by the system to reach the steady state. 
It has been observed in some earlier studies \cite{estz} that it also shows a scaling behaviour with the 
system size with the dynamical exponent $z$. This  in fact provides an alternative method to estimate $z$ 
when straight forward methods fail.

Both magntisation and energy are regarded as dimensionless quantities ($\epsilon$ and $E$ scaled by $J$) in this paper.

\end{enumerate}

\section{Detailed results of quenching dynamics on  RMA} 

The results of a zero temperature quench for the Ising model on the RMA are presented in this section.
Starting from a initial random configuration following a quench to zero temperature 
the system cannot reach the ground state (the state with zero residual energy) always
for any $p\neq0$. The magnetization, energy, persistence all attain a saturation value
in time. 
The saturation values of all the quantities show  nonmonotonic behaviour as a function of $p$.

\begin{figure}[h]
\rotatebox{0}{\resizebox*{8.8cm}{!}{\includegraphics{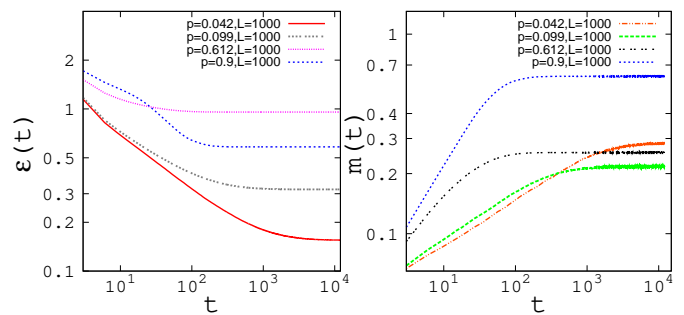}}}
\caption{(Color online) Decay of residual energy per spin and the growth of magnetisation with time for RMA for different probabilities.}  
\label{energy}
\end{figure}

Figure \ref{energy} shows the decay of residual energy per spin and the growth of magnetisation with time for different values of
the rewiring probability.

It is to be noted that the dynamic quantities do not show any obvious power law behaviour beyond a few time steps. 
For small $p$, there is apparently a power law behaviour for a larger  range of time which we believe is the effect of the
$p=0$ point where such a scaling definitely exists.

\begin{figure}[h]
\rotatebox{270}{\resizebox*{5.5cm}{!}{\includegraphics{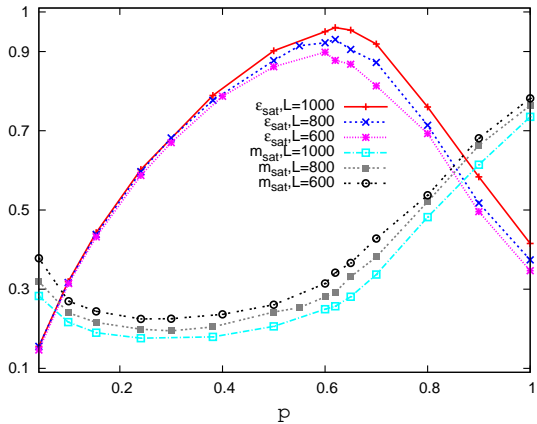}}}
\caption{(Color online) Saturation value of residual energy per spin $\varepsilon_{sat}$ and the saturation value of 
magnetisation $m_{sat}$ is plotted with the probability of rewiring $p$
for Random Model A.}  
\label{S_eng_mag}
\end{figure}

The saturation value of the residual energy per spin $\varepsilon_{sat}$ increases with the rewiring probability $p$ (for small $p$), reaches a maximum 
for an intermediate value of $p$ ($p <1 $) and then decreases again. This implies that the disorder 
of the spin system is maximum for a non trivial value of $p=p_{maxdis}$, which can be termed as the point of maximum disorder. 
The saturation value of magnetization on the other hand decreases for small $p$
and takes its minimum value for another intermediate value of $p$ ($p <1 $), and then increases again (Fig. \ref{S_eng_mag}).

\begin{figure}[h]
\rotatebox{0}{\resizebox*{6.5cm}{!}{\includegraphics{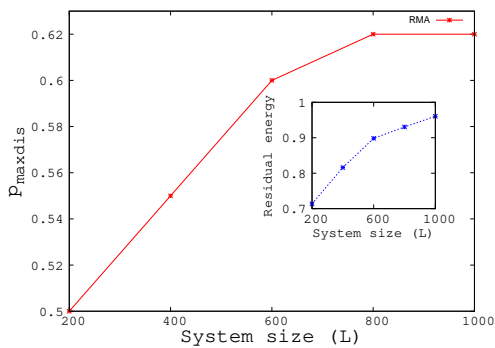}}}
\caption{(Color online) Rewiring probability at the point of maximum disorder is plotted with the system size.
Inset shows the increase of residual energy at the point of maximum disorder with the increment of the system size.}  
\label{wsensatp}
\end{figure}

$p_{maxdis}$ increases with the system size $L$ for small $L$ 
and then appears to saturate for larger system sizes. The value of the residual energy at $p_{maxdis}$ also increases 
with the system size (Fig. \ref{wsensatp}). This establishes  the existence of the point of maximum disorder at an 
intermediate value of $p$ ($p \simeq 0.62$) even in the thermodynamic limit.

Magnetisation reaches a minimum  at a value of $p$ which is {\it less}  than $p_{maxdis}$. This implies that there exists a
region where both magnetisation  and energy increase as $p$ increases. This is also apparent from Fig \ref{S_eng_mag}.
The physical phenomena responsible for  this intriguing feature is conjectured and discussed  in detail in sec VI B.

\begin{figure}[h]
\rotatebox{0}{\resizebox*{6.5cm}{!}{\includegraphics{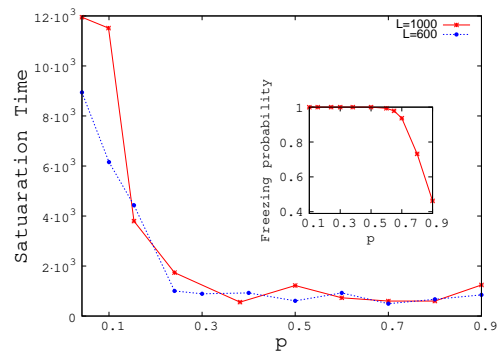}}}
\caption{(Color online) Time of saturation with the probability of rewiring is plotted for two different sizes for Random Model A.
Inset shows the variation of freezing probability with the probability of rewiring for RMA.}  
\label{frez}
\end{figure}

The saturation time decreases very fast with the rewiring probability $p$ for small $p$ and remains almost constant 
as $p$ increases (Fig. \ref{frez}). 
It is known that for $p=0$ the saturation time varies as $L^2$, here it appears that for any $p> 0$, there is
no  noticeable size dependence. 

For RMA, the freezing probability is almost unity for small $p$.
However, when the disorder is increased beyond $p\simeq0.5$, the  freezing probability shows a rapid decrease (Fig: \ref{frez}, inset).
In one dimension, we checked that the freezing probability is $zero$ for the regular network ($p=0$), but here we find that 
even for very small values of $p$, the freezing probability is unity. So there is a   discontinuty in the  freezing probability at $p=0$.
This also supports the fact that any finite $p$ can make the dynamics different from a conventional coarsening process.

An interesting observation  may be made about the behaviour of the saturation value of the residual energy in the region $p < 0.5$.
If one allows $p$ to decrease from $0.5$ to $0$,  the saturation value of the residual energy also   decreases  although  the freezing probability is
unity in the entire region. This implies that in this range of the parameter,
 although the system does not reach the real ground state in any realisation of the 
network (or initial configuration), such that
$\epsilon \neq 0$ in each case, the system has a tendency to approach the    
 the actual ground state monotonically with $p$ for $p<0.5$ (Fig. \ref{S_eng_mag}).

\begin{figure}[h]
\rotatebox{0}{\resizebox*{8.1cm}{!}{\includegraphics{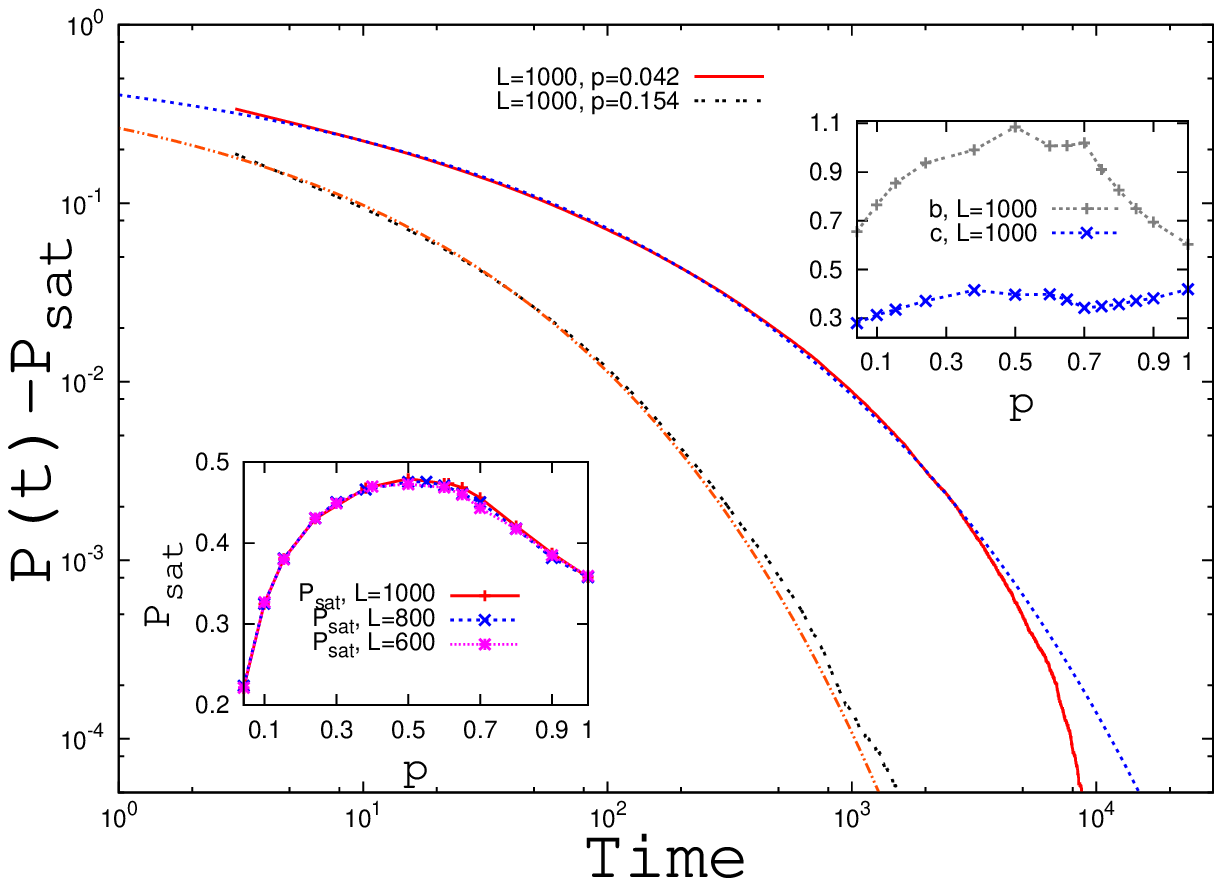}}}
\caption{(Color online) Decay of $P(t)-P_{sat}$ with time $t$ alongwith the stretched exponential function found to fit its form are shown. The inset in the bottom left 
shows the variation of the saturation value of persistence $P_{sat}$ with $p$. The other inset on the top right shows 
the variation of $b$ and $c$ with $p$.}  
\label{per_ws}
\end{figure}

The persistence probability follows a  stretched exponential behaviour  with time for any non zero $p$,
fitting quite well to the form
\begin{equation}
 P(t) - P_{sat} \simeq a \exp(-bt^c). 
\end{equation}
The saturation value of the persistence is $P_{sat}$, and it does not depend on the system size.
 $P_{sat}$ changes with the rewiring probability $p$ and there also exists an intermediate value of $p$ 
where the value of $P_{sat}$ is maximum. 
$b$ and $c$ vary nonmonotonically with $p$ (Fig. \ref{per_ws}).

\section{Detailed results of quenching dynamics on  RMB} 

\begin{figure}[h]
\rotatebox{0}{\resizebox*{8.8cm}{!}{\includegraphics{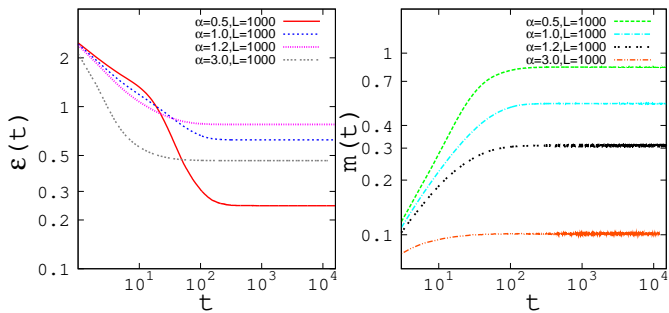}}}
\caption{(Color online) Decay of residual energy per spin and the growth of magnetisation with time for RMB for different probabilities.}  
\label{energy_eu}
\end{figure}

In this section we will present the results of the zero temperature quenching dynamics of Ising model on RMB.
Here also the system does not reach the ground state 
 always
for any finite value of $\alpha$. The magnetization, energy, persistence all attain a saturation value
in time as in RMA. Figure \ref{energy_eu} shows the decay of residual energy per spin and the growth of magnetisation with time 
for different values of $\alpha$. It is to be noted that the dynamical quantities do not show any obvious 
power law behaviour also for RMB.
 

\begin{figure}[h]
\rotatebox{270}{\resizebox*{5.5cm}{!}{\includegraphics{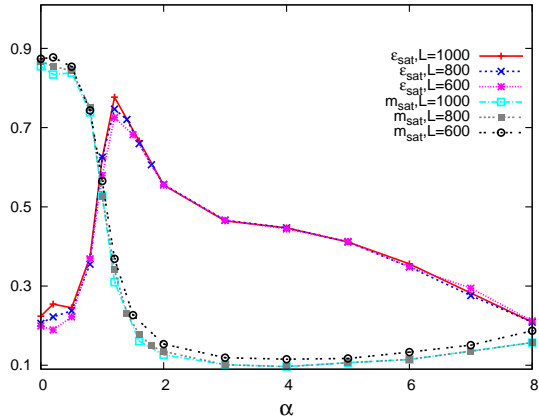}}}
\caption{(Color online) Saturation value of residual energy per spin $\varepsilon_{sat}$ and the saturation value of 
magnetisation $m_{sat}$ is plotted with $\alpha$ for Random Model B.}  
\label{S_eng_mag_eu}
\end{figure}

The saturation values of all the quantities show  nonmonotonic behaviour as a function of $\alpha$.
The saturation value of residual energy per spin $\varepsilon_{sat}$ increases with $\alpha$ for small $\alpha$, reaches a maximum 
for a finite value of $\alpha$ and then decreases again. This implies that for the RMB also, the disorder 
of the spin system is maximum for a finite value of $\alpha$, which is the point of maximum disorder here.
On the other hand, the saturation value of the magnetization  decreases for small $\alpha$
and takes its minimum value for another finite value of $\alpha$ and then slowly increases (Fig. \ref{S_eng_mag_eu}).

\begin{figure}[h]
\rotatebox{0}{\resizebox*{6.5cm}{!}{\includegraphics{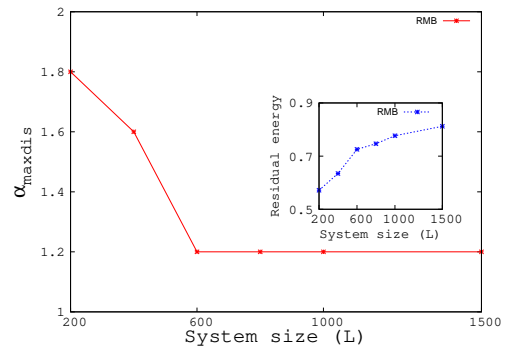}}}
\caption{(Color online) The value of $\alpha$ at the point of maximum disorder is plotted with the system size.
Inset shows the increase of residual energy at the point of maximum disorder with the increment of the system size.}  
\label{euensatp}
\end{figure}

The value of $\alpha= \alpha_{maxdis}$, at which the  maximum disorder occurs, decreases with the system size $L$ for small $L$ 
and then saturates for larger system sizes. The value of the residual energy at  $\alpha_{maxdis}$ also increases 
with the system size (Fig. \ref{euensatp}). This establishes  the existence of the point of maximum disorder, for the 
RMB, at a finite value of $\alpha$ ($\alpha \simeq 1.2$) even in the thermodynamic limit.
Similar to the RMA, here  is a region beyond $\alpha = 1.2$ where the energy and the magnetisation both decrease,
until the magnetisation starts growing again. As already mentioned, this issue is addressed in sec. IV B.

Saturation time for RMB in the random network in the region $0\leq \alpha <1$ shows too large fluctuations to let one conclude 
whether it is a constant in this region or has a variation with $\alpha$. Beyond $\alpha =1$ and upto $\alpha =3.0$, it is 
almost independent of $\alpha$. For $\alpha > 3 $ 
the saturation time increases with $\alpha$. There is no remarkable finite size effect in the saturation time for the 
RMB for any finite value of $\alpha$. The saturation time 
varies as $L^2$ for a regular lattice corresponding to $\alpha \to \infty$, here it appears that for any finite $\alpha$, however large, there is no  remarkable size dependence. 

\begin{figure}[h]
\rotatebox{0}{\resizebox*{6.7cm}{!}{\includegraphics{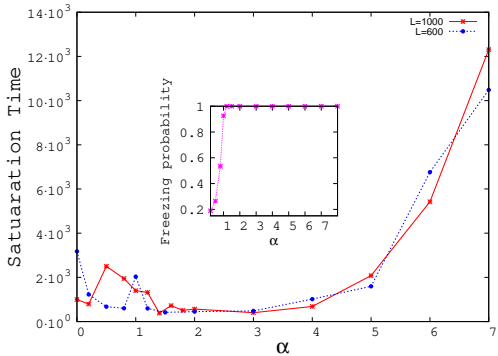}}}
\caption{(Color online) Time of saturation with the value of $\alpha$ is plotted for two different sizes for Random Model B.
Inset shows the variation of freezing probability with $\alpha$ for RMB.}  
\label{frez_eu}
\end{figure}

The freezing probability is small for $\alpha =0$ ($\simeq 0.2$) and increases rapidly with $\alpha$ for small $\alpha$.
Freezing probability becomes almost unity beyond $\alpha \simeq 1.2$ and remains the same for large $\alpha$. It seems that 
for any finite  $\alpha > 1.2$ freezing probability remains unity and it will be zero only at  $\alpha \rightarrow \infty$ 
(Fig. \ref{frez_eu}), as in one dimension, the freezing probability is $zero$ for the regular network. 
So for RMB there is a  discontinuity of freezing probability at $\alpha = \infty$ which corresponds to the $p=0$ point of RMA. 

Beyond $\alpha \simeq 1.2$, the energy  decreases with $\alpha$ though the freezing probability remains
unity. 
This implies that although the system definitely reaches a frozen state, it approaches  the 
real ground state monotonically as $\alpha \to \infty $  (Fig. \ref{S_eng_mag_eu}).

The above results indicate that, though for $\alpha >2 $ the network behaves as a regular one,  dynamically the network is regular 
only at its extreme value $\alpha \rightarrow \infty$.

\begin{figure}[h]
\rotatebox{0}{\resizebox*{8.2cm}{!}{\includegraphics{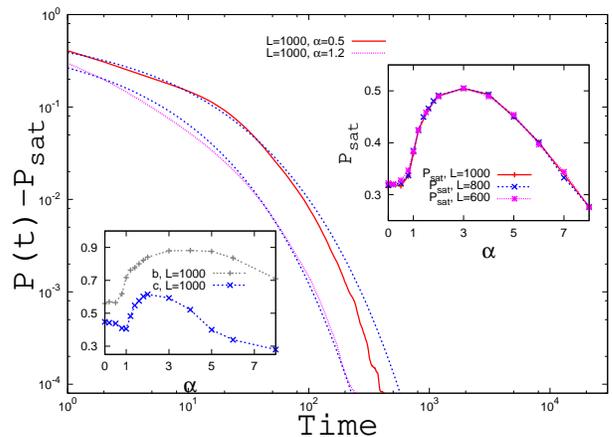}}}
\caption{(Color online) Decay of $P(t)-P_{sat}$ with time $t$ along with   the stretched exponential function found to fit it approximately  are shown. The inset in the top right 
shows the variation of the saturation value of persistence $P_{sat}$ with $\alpha$. The other inset on the bottom left shows 
the variation of $b$ and $c$ with $\alpha$.}  
\label{per_eu}
\end{figure}

We find that the persistence probability follows roughly a stretched exponential form with time (given by equation (2)) 
for any finite $\alpha$. 
The saturation value of the persistence, $P_{sat}$, does not depend on the system size.
$P_{sat}$ changes with $\alpha$ and there exists an intermediate value of $\alpha$ 
where the value of $P_{sat}$ is maximum. 
For RMB also $b$ and $c$ vary nonmonotonically with $\alpha$ (Fig: \ref{per_eu}). 

\section{Discussions on the results}

\subsection{Comparison of the results for RMA and RMB} 

In the last two sections the results of a quench at zero temperature for the Ising model on  RMA and RMB have been presented separately.
In this subsection we will compare the results  to understand  how the difference in the nature of randomness 
affects the dynamics of Ising spin system.

The gross features of the results are similar: in both models we have a freezing effect which makes the system
get stuck in a higher energy state compared to the static equilibrium state in which all spins are parallel.
No  power law scaling  behaviour with time is observed in the dynamic quantities in either model. There exists 
a point in the parameter space where the deviation from the static ground state is maximum. The 
behaviour of the saturation times and freezing  probability as functions of the disorder parameters are also
quite similar qualitatively.

The saturation values of magnetisation
 and persistence  attain a minimum and maximum value respectively at an intermediate value of the relevant parameters in both models.
The decay of the persistence probability also follows the same functional form in the  entire parameter 
space.  The saturation values of the persistence has no size dependence for both the models.
This indicates that as a whole the dynamics is not much affected due to the change in the nature of randomness of the Ising spin system.

Let us  consider  the parameter values at which the   RMA and RMB are equivalent as a network: 
RMA and RMB  behave as random networks at $p=1$ and  $\alpha =0$ respectively. So one can expect that the saturation values of 
residual energy per spin, magnetisation and the numerical value of the saturation time would be same at these values.
However, the numerical values of these quantities are quite different. 
For RMA, at $p=1$ the saturation value of the residual energy per spin $\varepsilon_{sat} \simeq 0.415$  
whereas for RMB at  $\alpha =0$ $\varepsilon_{sat} \simeq 0.224$ for $L=1000$. Similarly we found numerically that 
for RMA the value of saturation magnetisation  $m_{sat} \simeq 0.735$ for RMA and $m_{sat} \simeq 0.855$ for RMB for the same system size.
This is because even though the networks are both random here, the connections have a subtle difference.  
For RMA, the number of second nearest neighbour is exactly zero at 
$p=1$ and all the other long range neighbour connections are equally probable. On the other hand, 
for RMB, second nearest neighbours can be still present in 
the network and the probability is same for this  and any other longer range connection. This difference 
in the nature of randomness affects the dynamics of the Ising spin system sufficiently to make the saturation values different. This 
means that the systems are locked at {\it different} nonequilibrium steady states.
For RMB, it is closer to the actual ground state as it is more short ranged in comparison.

The other values at which the two networks are equivalent are  $p=0$ and $\alpha > 2$ where regular network behaviour 
is found as far as the network properties are concerned. Interestingly, the behaviour of RMB even 
when $\alpha$ is finite and greater than 2, is not quite like the dynamics of a regular one dimensional lattice with  
nearest and next nearest neighbour links only. In fact, the point at which the magnetisation becomes minimum is well inside 
the region $\alpha > 2$ and not within the small world region as in RMA. Actually there is an extended region of regular and 
random network behaviour for the RMB, and as a result, a few more interesting points are
possible to observe  here.  Only at the extreme point $\alpha = \infty$, the 
one dimensional Ising exponents $z = 2.0$ and $\theta = 0.375$ can be recovered as the frozen states continue to exist 
even for finite values of  $\alpha > 2$ for RMB. 
For the regular network with nearest and next nearest neighbour model,
we have checked that there is no freezing at all. So discontinuities of the freezing probabilities occur at $p=0$ 
and $\alpha = \infty$ on RMA and RMA respectively. 

Though the nature of randomness is different for RMA and RMB, for both the models there exists a point of maximum disorder where 
the saturation value of the residual energy per spin attains a maximum value. 
For RMB, maximum disorder of the Ising spin system occurs near the static phase transition point (small 
world to random phase) whereas for RMA, the point of maximum disorder is well within the small world region.

We try to explain this considering the deviation from the point $p=1$ (for RMA) and $\alpha = 0$ (for RMB). 
Two processes occur simultaneously here:
(a) Number of connections with further neighbours decreases and (b) clustering becomes more probable. 
As a result of these two processes,  freezing occurs. For RMA, the effect is $less$ as there is 
less clustering \cite{gbs}. But for RMB, the effect is $more$ and spans the entire parameter space $\alpha >1$ and 
therefore the point of maximum disorder of Ising spin system is very close to the random - small world phase transition point
$\alpha =1$.

The question may arise whether this difference prevails when the models are made even more similar.
In RMB, the probability $p_3(\alpha)$ that $l \geq 3$ can be expressed as a function of $\alpha$:  
 \begin{equation}
 p_3(\alpha)= \frac{\sum_{l=3}^{l=L/2} l^{-\alpha}}{\sum_{l=2}^{l=L/2} l^{-\alpha}} .
 \end{equation}
 A further correspondence between the two networks can be established by imposing  
  $p=p_3(\alpha)$,  which makes the number of second neighbour links 
 in RMA and RMB also same  (but the rest of the extra links are connected differently). 

Using equation (3) we can obtain the value of $p$ corresponding to a given  value of $\alpha$ and vice versa.
But it is immediately seen that the two networks are not equivalent even after making them similar upto
the second neighbour connections. For example, for $\alpha = 2.0$, the corresponding value of $p =0.612$ in this 
scheme. But we have already seen that while the point of maximum disorder occurs close to  this value of $p$ in RMA, 
the point of maximum disorder  for RMB is considerably away from  $\alpha = 2.0$.
So the nature of randomness continues to affect the dynamics at least quantitatively.

\subsection{Analysis of some general features of the quenching  phenomena on networks}

We find several interesting features in the quenching phenomena of Ising spin systems on both the networks and  in this
subsection we attempt to provide an understanding of the same.

It is intriguing that the results indicate that the minimum amount of randomness can 
make the system freeze. What happens for small randomness? The interactions are still dominantly 
nearest neighbour type and domains in the conventional sense should grow which will be of both plus and minus signs.
The system will freeze as there will be some stable domain walls due to the few long range  interactions present.
The domains, as the system attains saturation, will be small in number and large in size irrespective of their signs.
As a result, the magnetisation attains a small value while the residual energy is still small.

This effect continues for some time till something more interesting happens. Take for example the 
case of quenching on RMA. There is a distinct region  $0.4 < p < 0.6$ where the energy and 
magnetisation grow simultaneously, an apparently counterintuitive result. Similar behaviour can be noted for the quenching on RMB in   a certain 
region in its 
parameter space. A problem to analyse the situation for different $p$ (or $\alpha$) values 
is that the final frozen states are not related in any way in principle. This is because the energy landscapes change as $p$
is changed and the initial configurations which undergo evolution are completely uncorrelated. 
In fact, in such a situation, even if the energy landscape is same with a number of local minima, 
different initial configurations will end up in different final nonequilibrium steady states.
 Nevertheless, one can attempt to 
explain this  counterintuitive result assuming that the final states are not largely 
different when $p$ is changed slightly in the following way. This assumption and explanation  are
supported by the actual final states obtained for small system sizes.

\begin{figure}[h]
\rotatebox{270}{\resizebox*{5.5cm}{!}{\includegraphics{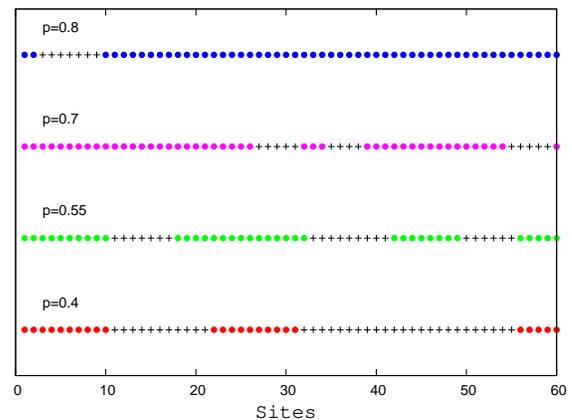}}}
\caption{(Color online) Snap shots of the final spin configurations for different values of the disorder parameter $p$ for quenching on RMA.   
The $+$ and $\bullet$ signs indicate up and down spins respectively. The domains  in the conventional sense are clearly visible.} 
\label{snap}
\end{figure}

Let us for example consider the RMA and take two values of $p$, $p_2>p_1$, and for which  the magnetisation 
and residual energy of the final state corresponding $p_2$ are both larger than those for $p_1$. 
Now this can  be possible due to the fragmentation of a larger domain into several 
domains such that the magnetisation increases. This can be demonstrated with a simple example: 
let us imagine a situation where one has only two domains of size $N^+$ (of up spins) and $N^-$ 
(of down spins) for $p_1$ with magnetisation equal to $ m_1 = |(N^+ - N^-)|/L$ and assume 
that for $p_2$, the domain with $N^+$ up spins remains same while the domain with $N^-$ down spins gets 
fragmented into three domains  of size $N^-{_1}, N{^+}{_1} $ and $ N^-{_2}$ in the final state. For $p_2$, 
therefore,  the magnetisation is $m_2 = |(N^+ + 2N^+{_1} - N^-)|/L$  which is  larger than $m_1$. 
Here in this hypothetical case, we have assumed that $N^+ >N^-$, and $p_2$ is very close to $p_1$. 
One can also assume that the energy increases for $p_2$ as the system is still sufficiently short ranged and 
the new domain walls cause an extra energy compared to the state obtained for $p_1$.
Of course this is an oversimplified picture where we have assumed that the final states for $p_1$ and $p_2$ 
are identical except for the fragmentation of one domain. However, we find that the final configurations obtained for small systems
for different values of $p$ as shown in Fig. \ref{snap} are consistent with our conjecture. These snapshots 
are representative of the real situation in the sense that they give a  typical picture and are not  just  
rare  cases; we have obtained a similar picture from almost all such configurations generated for small systems.

As $p$ further increases, should the domains get fragmented into even smaller pieces? Answer is no, as
the increasing number of long range interactions again help in the growth of so called domains, {\it of one particular
sign only} such that the magnetisation grows and the energy decreases. However, domains of both signs still survive, 
although the sizes are no longer comparable. It can therefore be expected that the region for which both magnetisation and energy increase 
as a function of $p$ or $\alpha$ would continue till the short range interactions are dominating and our results
are consistent with this expectation.

\section{Summary and concluding remarks} 

In this paper, we addressed the question how  the quenching dynamics 
of Ising spins depend on the nature of randomness  of the underlying network by considering two networks in which the
randomness is realised differently. 
The networks are same upto the first neighbour
links and have same average degree per node. While the qualitative features are 
same, there are intricate differences occurring in the behaviour of the saturation values of the dynamical quantities.

Overall, we find some interesting features: the saturation values of the dynamical quantities do not 
have monotonic behaviour as a function of the disorder parameters. Especially, we find that increasing 
randomness does not necessarily make the system get locked in a higher energy state. The dynamics
takes the system to a steady state very fast, and the saturation times are not dependent on the system size.
No scaling behaviour is obtained from the studies either with time or with system size
for any of the dynamic quantities.
The most surprising result is perhaps the existence of a region in the parameter space where 
both the residual energy and the magnetisation increase which can be explained phenomenologically.

The Euclidean model, on which the study of the  
quenching of Ising spins is done for the first time to the best of our knowledge,
shows some surprising behaviour both in the random and regular regions. We find that
 decreasing randomness makes the system end up in a higher energy state in the random region while
in the regular region, familiar behaviour of the Ising dynamics with short range interactions
are not obtained; in fact the  probability of freezing  is unity here indicating that
in none of the realisation, the system could end up in the static ground state. 
The saturation time also does not show scaling with time.

As already mentioned, the present study is relevant for dynamical 
social phenomena on complex networks. For example, 
the evolution of binary  opinions on a complex network 
(where the initial states are randomly $+1$ and $-1$) 
 is analogous to the dynamical 
study reported in the present paper. Of course, in case of the opinion dynamics,
the interactions could be more complex compared to the  the simple Ising model.  Our result indicates that the qualitative features of the results will  not be 
 much different for different complex networks.

Dynamic frustration \cite{pratap_ps} is responsible for freezing in many Ising systems where 
there is no frustration in the conventional sense. 
One interesting observation is that the nature of 
dynamic frustration in regular lattices of dimension greater than one and that in 
systems with random interaction (but no frustration) are in general quite different as in the
latter one does not encounter the familiar scaling laws.

Acknowledgements :

Financial support from DST project SR/S2/CMP-56/2007 and partial computational support from UPE project are
acknowledged (SB and PS). SB acknowledges financial support from CSIR (Grant no. 09/028(0818)/2010-EMR-I).

\end{document}